\documentclass{jfm}
\usepackage{graphicx}
\usepackage{epstopdf, epsfig}
\usepackage[usenames, dvipsnames]{color}

\shorttitle{The origins of torque enhancement through roughness in TC}
\shortauthor{X. Zhu, R. Verzicco and D. Lohse}

\title{Disentangling the origins of torque enhancement through wall roughness in Taylor-Couette turbulence}

\author{Xiaojue Zhu\aff{1}
 Roberto Verzicco\aff{2,1}
 \and Detlef Lohse\aff{1,3}}

\affiliation{\aff{1}Physics of Fluids Group, MESA+ Institute and J. M. Burgers Centre for Fluid Dynamics, University of Twente, P.O. Box 217, 7500AE Enschede, The Netherlands
\aff{2}Dipartimento di Ingegneria Industriale, University of Rome "Tor Vergata",
Via del Politecnico 1, Roma 00133, Italy
\aff{3}Max Planck Institute for Dynamics and Self-Organization, 37077 G\"ottingen, Germany}

\begin{document}

\maketitle

\begin{abstract}
Direct numerical simulations (DNSs) are performed to analyze the global transport properties of turbulent Taylor-Couette flow with inner rough wall up to Taylor number $Ta=10^{10}$. The dimensionless torque $Nu_\omega$ shows an effective scaling of $Nu_\omega \propto Ta^{0.42\pm0.01}$, which is steeper than the ultimate regime effective scaling $Nu_\omega \propto Ta^{0.38}$ seen for smooth inner and outer walls. It is found that at the inner rough wall, the dominant contribution to the torque comes from the pressure forces on the radial faces of the rough elements; while viscous shear stresses on the rough surfaces contribute little to $Nu_\omega$. Thus, the log layer close to the rough wall depends on the roughness length scale, rather than on the viscous length scale. We then separate the torque contributed from the smooth inner wall and the rough outer wall. It is found that the smooth wall torque scaling follows $Nu_s \propto Ta_s^{0.38\pm0.01}$, in excellent agreement with the case where both walls are smooth. In contrast,  the rough wall torque scaling follows $Nu_r \propto Ta_r^{0.47\pm0.03}$, very close to the pure ultimate regime scaling $Nu_\omega \propto Ta^{1/2}$. The energy dissipation rate at the wall of inner rough cylinder decreases significantly as a consequence of the wall shear stress reduction caused by the flow separation at the rough elements. On the other hand, the latter shed vortices in the bulk that are transported towards the outer cylinder and dissipated. Compared to the purely smooth case, the inner wall roughness renders the system more bulk dominated and thus increases the effective scaling exponent.
\end{abstract}


\section{Introduction}
Taylor-Couette (TC) flow \citep*{grossmann2016}, in which a fluid is confined between two rotating cylinders, and Rayleigh-B\'enard (RB) flow \citep*{ahlers2009}, in which a fluid is heated from below and cooled from above, are the two most famous examples for turbulent flows in closed system, in which exact energy balances hold and global transport properties can be connected to the energy dissipation rates. Due to the close analogy \citep*{eckhardt2007a,eckhardt2007b}, they have been called the twins of turbulence research \citep{busse2012}. In the TC system, the global transport property is often expressed as torque $\tau$, which drives the cylinder at a constant angular velocity. Following the analogy \citep*{eckhardt2007b} between the angular velocity flux from the inner to the outer wall in TC and heat flux from bottom to top plate in RB flow, the torque in dimensionless form can be expressed as a Nusselt number $Nu_\omega$, which for smooth cylinders and in the ultimate turbulent regime was shown to have an effective scaling $Nu_\omega \propto Ta^{0.38}$ with the later defined Taylor number $Ta$ \citep{vangils2011,ostilla2014a,ostilla2014b}, identical to the effective scaling $Nu \propto Ra^{0.38}$ found in ultimate RB turbulence in terms of Nusselt number $Nu$ and Rayleigh number $Ra$ \citep{he2012a,he2012b}. This effective scaling in the ultimate regime has been explained by \cite{grossmann2011} as the pure ultimate regime scaling exponent 1/2 \citep{kraichnan1962} with logarithmic correction due to the interplay between bulk and turbulent boundary layer. 

Altering the boundary conditions by implementing rough boundaries is thought to be one way to reduce the logarithmic correction because roughness can modify the boundary layer and hence the boundary-bulk interaction. In RB convection, roughness has already been employed in many studies \citep{shen1996,du2000,ciliberto1999,stringano2006,roche2001,wagner2014}, however, the results concerning the change of the scaling exponent are still elusive \citep{ahlers2009,tisserand2011}. For TC flow the studies regarding rough boundaries have attracted much less attention. The only two examples are by \cite{cadot1997} and \cite{berg2003} and indeed for rough walls an increase of the torque scaling exponent towards the pure ultimate scaling $Nu_\omega \propto Ta^{1/2}$ has been found. Those global measurements however did not provide information on \textit{local} flow details, therefore the exact mechanism of the torque enhancement could not be elucidated. Most importantly, knowledge on how the roughness changes the flow structure and how the flow structure affects the torque scaling is still missing.

In this manuscript, direct numerical simulations (DNS) are performed to simulate Taylor-Couette flow with a rough \textit{inner} wall. This configuration allows us to dissect directly smooth turbulent boundary layer, rough turbulent boundary layer, and bulk interactions simultaneously. Our aim here is to demonstrate that, a) pressure dominance of the torque at the rough wall is of prime importance in changing the flow structure; b) pressure induced separation causes an increase of the dissipation through the vortex-shedding towards the bulk, rather than via a large shear rate very close to the wall as in the smooth case. The consequence is an increase of the scaling exponent.

\section{Numerical techniques and parameters}\label{sec:rules_submission}
\begin{figure}
\centering
\includegraphics[width=3.0in]{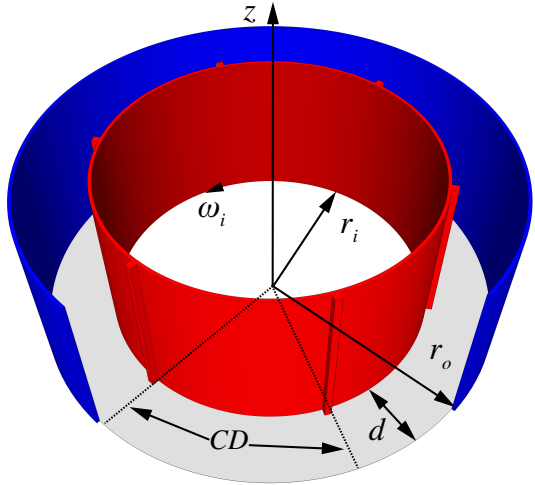}
 \caption{Schematic view of the Taylor-Couette system with roughness. The inner cylinder with radius $r_i$ is rotating with angular velocity $\omega_i$. The outer cylinder with radius $r_o$ is at rest. The gap width is $d=r_o-r_i$. The rough elements are six square vertical stripes positioned equidistantly on the inner cylinder wall. The height of the rough element is $0.1d$. In our DNS, a rotational symmetry of six is used. The computational domain contains 1/6 of the azimuthal width and has one rough element on the inner cylinder. CD: computation domain.}
\label{figs}
\end{figure}

For the DNS a second-order finite difference code \citep{verzicco1996,vandepoel2014} is employed, in combination with an immersed boundary method \citep{fadlun2000} to deal with the roughness. The radius ratio is $\eta=r_i/r_o=0.714$, where $r_i$ and $r_o$ are the inner and outer radii, respectively. The aspect ratio of the computational domain is $\Gamma=L/d=2.094$, where $L$ is the axial periodicity length and $d$ the gap width $d=r_o-r_i$. With such $\Gamma$, we can have a relatively small computational box with a pair of Taylor vortices of opposite directions. The inner cylinder is roughened
by attaching six vertical strips of square cross section
(edge width $h=0.1d$) which are equally spaced in azimuthal
angle, similar to the procedure used in \citep{cadot1997,berg2003}. A schematic view of the geometry is seen in figure \ref{figs}. A rotational symmetry of order six is implemented to reduce computational cost while not affecting the results \citep{braukmann2013,ostilla2014a}. As a result, in our azimuthally reduced domain, only one square rough element exists. Here we focus on the case of inner cylinder rotation with angular velocity $\omega_i$ and fixed outer cylinder. The appropriate number of grid points is chosen to make sure that the torque difference between the inner and outer cylinder is less than one percent. The parameters are shown in detail in table \ref{tab1}.

In order to show the modification of roughness to the global transport properties, from DNSs, we extract $Nu_\omega$ and the wind Reynolds number $Re_w$ as a function of $Ta$, which are defined as, 
\begin{eqnarray}\label{EQ1}
Nu_\omega = \tau/\tau_{pa}, 
\end{eqnarray}
where $\tau_{pa}$ is the torque required to drive the system in the purely azimuthal and laminar case, and
\begin{eqnarray}\label{EQ2}
Re_w=\sigma(u_r)d/\nu, 
\end{eqnarray}
where $\sigma(u_r)$ is the standard deviation of radial velocity and $\nu$ the kinematic viscosity of the fluid. The Taylor number is defined as
\begin{eqnarray}\label{EQ3}
Ta=\frac{1}{64} \frac{(1+\eta)^4}{\eta^2}d^2{(r_i+r_o)}^2\omega_i^2{\nu}^{-2},
\end{eqnarray}
where $\omega_i$ is the angular velocity of the inner cylinder. Another alternative way to characterize the system is by using the inner cylinder Reynolds number $Re_i=Ud/\nu=r_i \omega_i d/ \nu$ \citep{lathrop1992b,lewis1999}. Note that these two definitions can be easily transformed between each other by the relation
\begin{eqnarray}
Ta=\frac{(1+\eta)^6}{64\eta^4}Re_i^2.
\end{eqnarray}
The torque $\tau$ can be further related to the friction velocity $u_\tau$ and the viscous length scale $\delta_\nu$:

\begin{eqnarray}\label{EQ4}
u_\tau=\sqrt  {\tau / 2\rho \pi r^2 L },
\end{eqnarray}
\begin{eqnarray}\label{EQ5}
\delta_\nu=\nu/u_\tau,
\end{eqnarray}
where $\rho$ is the density of fluid and $r$ can be either the inner cylinder radius $r_i$ or the outer one $r_o$. 

The torque can also be expressed in the form of a friction factor $C_f$, which is widely used for wall bounded turbulence. The relation between $C_f$ and $Nu_\omega$ is
\begin{eqnarray}
C_f=2\pi Nu_\omega J^\omega_0 \nu^{-2}/Re_i^2,
\end{eqnarray}
where $J^\omega_0=2\nu r_o^2 r_i^2 \omega_i/(r_o^2-r_i^2)$ is the angular velocity current at the purely azimuthal laminar state. 

 \begin{table}
\begin{center}
  \begin{tabular}{ccccccc}
   $Case No. $ & $Ta$ & $Re_i$ & $N_\theta \times N_r \times N_z$ & $Nu_\omega$ & $Re_w $ & $\eta_K/d$\\ [3pt]
   
1  & $9.75\times10^7$ & $8.00\times10^3$ & $512\times512\times512$ &  19.5 & $5.60\times10^2$ & $4.99\times10^{-3}$  \\
2  & $2.15\times10^8$ & $1.19\times10^4$ & $512\times512\times512$ &  26.0 & $8.63\times10^2$  &$3.80\times10^{-3} $\\
3  & $4.62\times10^8$ & $1.74\times10^4$ & $768\times640\times640$ &  35.9 & $1.25\times10^3$   &$2.88\times10^{-3}$\\
4  & $9.75\times10^8$ & $2.53\times10^4$ & $1024\times768\times768$ &  47.1 & $1.75\times10^3$  & $2.24\times10^{-3}$\\
5  & $2.15\times10^9$ & $3.76\times10^4$ & $1280\times1024\times1024$ &  65.5 & $2.71\times10^3$&  $1.69\times10^{-3}$ \\
6  & $4.63\times10^9$ & $5.52\times10^4$ & $1536\times1280\times1280$ &  91.2 & $3.95\times10^3$ & $1.28\times10^{-3}$\\
7  & $1.00\times10^{10}$ & $8.10\times10^4$ & $2048\times1536\times1536$ &  125.9 & $5.79\times10^3$&  $9.73\times10^{-4}$ \\

  \end{tabular}
  \caption{Values of the control parameters and the numerical results of the simulations. We keep the radius ratio at $\eta=0.714$ and the square rough element height $h=0.1d$, but we vary the $Ta$ and thus the $Re_i$ number. The fourth column shows the amount of grid points used in azimuthal ($N_\theta$), radial ($N_r$), and axial direction ($N_z$). The fifth column shows the dimensionless torque, $Nu_\omega$. The sixth column shows the wind Reynolds number $Re_w$. The last column details the mean Kolmogorov scale $\eta_K$. It is obtained from the relation $\eta_K=(\nu^3/\epsilon_{u,m})^{1/4}$, where $\epsilon_{u,m}$ is the mean kinetic dissipation rate Eq. (\ref{EQ14}). All of the simulations were run in reduced geometry with $L=2\pi/3$ and a rotation symmetry of the order 6. The corresponding cases at the same $Ta$ without roughness (with smooth cylinders) can be found in \cite{ostilla2013,ostilla2014a}.}
  \label{tab1}
  \end{center}
\end{table}

\section{Results}\label{sec:types_paper}
\subsection{Global scaling laws}
\begin{figure}
 \begin{minipage}[c]{0.5\textwidth}
    \includegraphics[width=2.8 in]{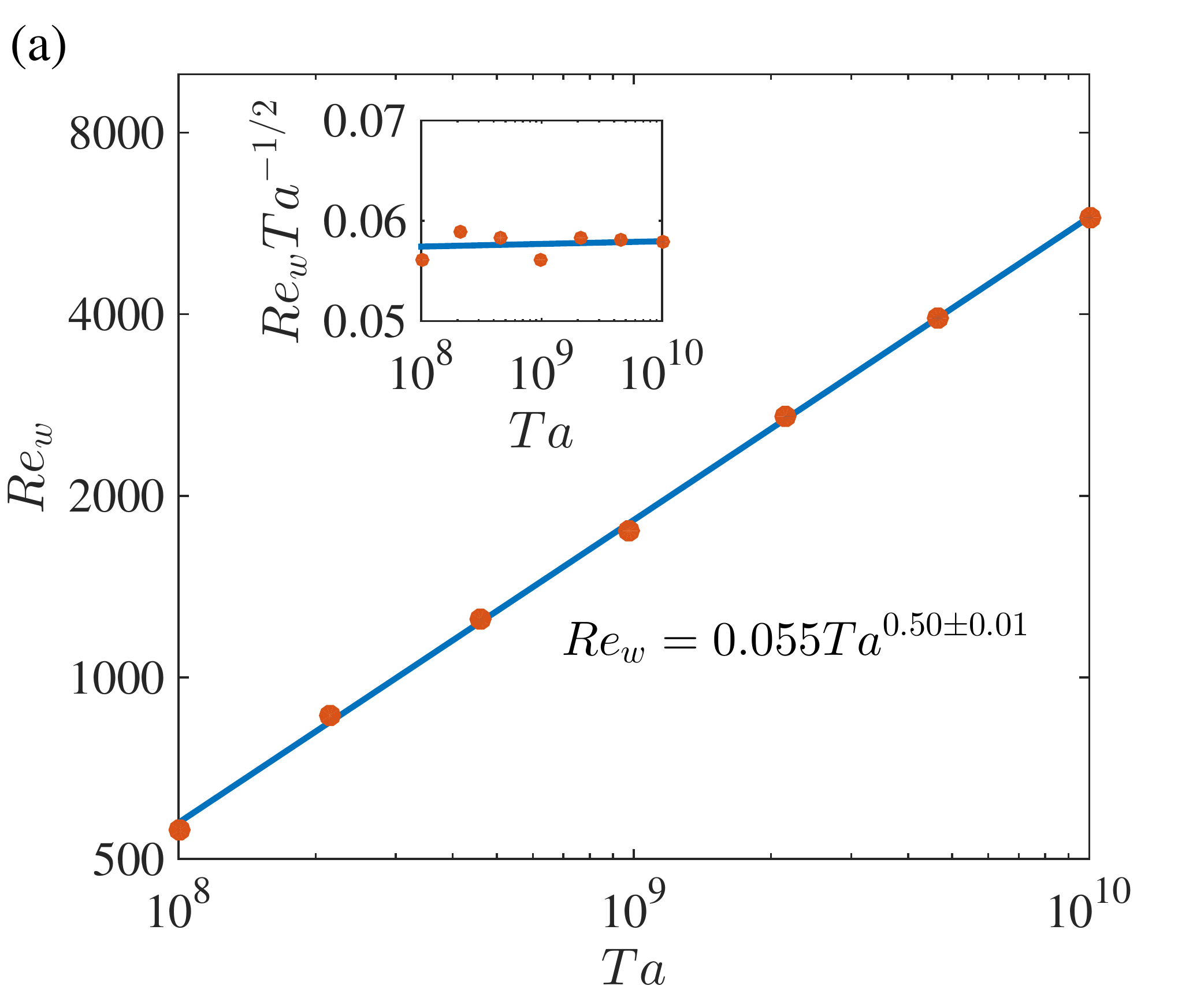}
   
  \end{minipage}%
     \begin{minipage}[c]{0.5\textwidth}
    \includegraphics[width=2.8 in]{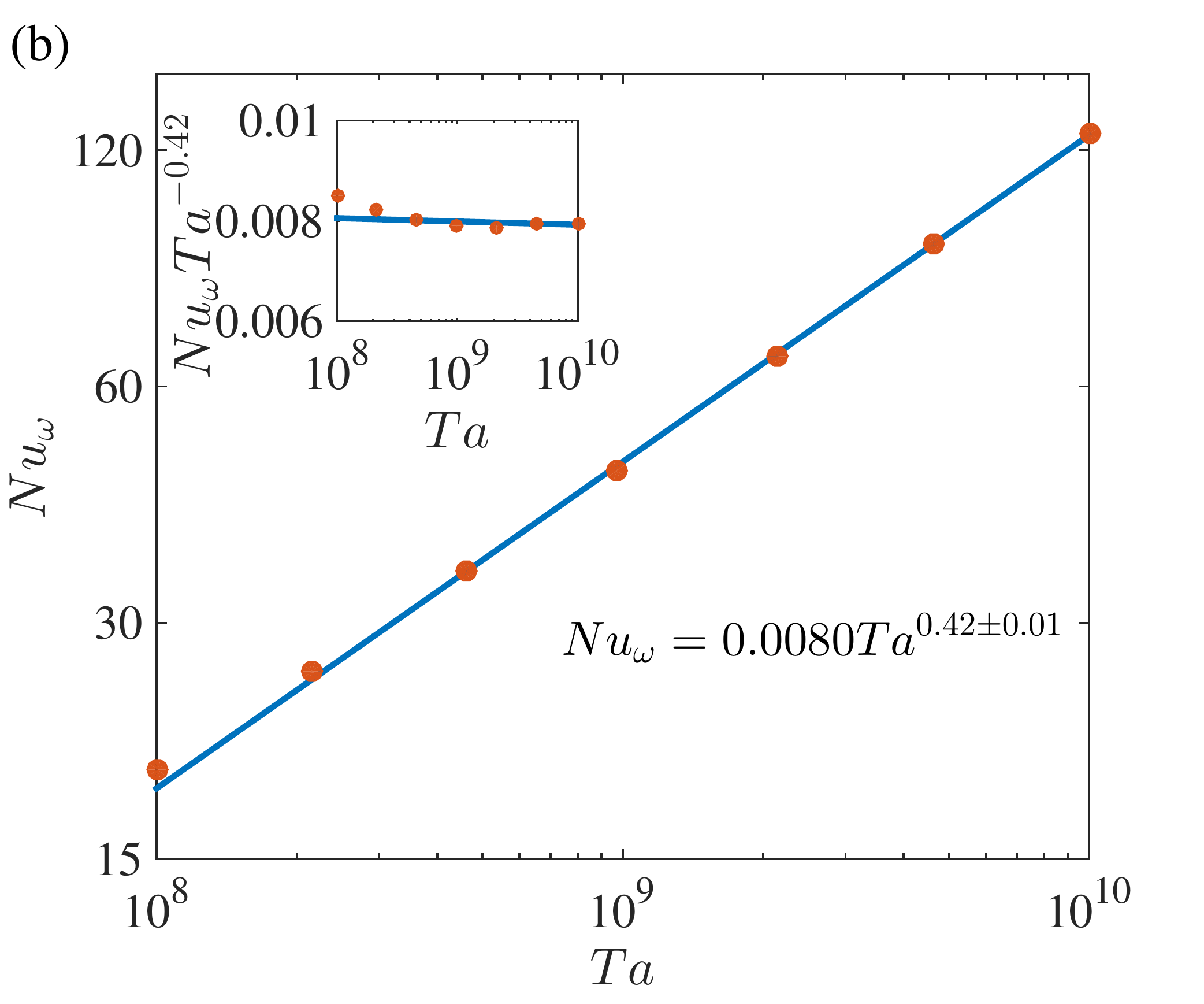}
    
  \end{minipage}
    \caption{(a) Wind Reynolds number $Re_w$ vs. Taylor number $Ta$. (b) Nusselt number $Nu_\omega$ vs. $Ta$. The bullets are the data from simulations while the straight lines are the best fits $Re_w=0.055Ta^{0.50\pm0.01}$ and $Nu_\omega=0.0080Ta^{0.42\pm0.01}$. The insets show the compensated plots $Re_w/Ta^{1/2}$ vs. $Ta$ and  $Nu_\omega/Ta^{0.42}$ vs. $Ta$. }
\label{fig1}
\end{figure}

The effective scaling relations of $Re_w$ and $Nu_\omega$ vs. $Ta$ are shown in figure. \ref{fig1}. We find $Re_w \propto Ta^{0.50\pm0.01}$, in accordance with the theory $Re_w \propto Ta^{1/2}$ \citep{grossmann2011} predicted for the ultimate regime with smooth boundary. This effective scaling was also found for ultimate RB \citep{he2012b} and ultimate TC \citep{vangils2011,huisman2012} turbulence. Indeed,
the prediction for $Re_w(Ta)$ of \cite{grossmann2011} should also work for
rough boundaries in the ultimate regime. It originates from the dominance of the turbulent bulk contribution to the energy dissipation rate. 
Once the BLs become even more turbulent through the roughness, the system will become more bulk-like and the scaling therefore should not change. The prefactor 0.055 in the effective scaling relation is larger than its counterpart for the smooth TC case where it is only 0.0424 \citep{huisman2012} which shows that roughness facilitates the fluctuations of the wind flow rather than hindering it.

We now come to the scaling of the torque. Here, the effective scaling relation $Nu_\omega \propto Ta^{0.42\pm0.01}$ is found. The scaling exponent is notably larger than the effective ultimate scaling exponent 0.38 seen for both smooth RB \citep{he2012b} and TC \citep{huisman2012} flows, respectively, in the appropriate $Ra$ and $Ta$ regimes. This effective scaling exponent originates from the pure ultimate scaling $Nu_\omega \propto Ta^{1/2}$ \citep{kraichnan1962}, but with logarithmic corrections \citep{grossmann2011}. Our finding of an increased effective scaling exponent as compared to that in the smooth case implies that roughness can \textit{reduce} the logarithmic correction. This becomes possible through redistribution of the energy dissipation in radial direction. We will explain the physics behind this statement in the following paragraphs.

\begin{figure}
\centering\includegraphics[width=2.8in]{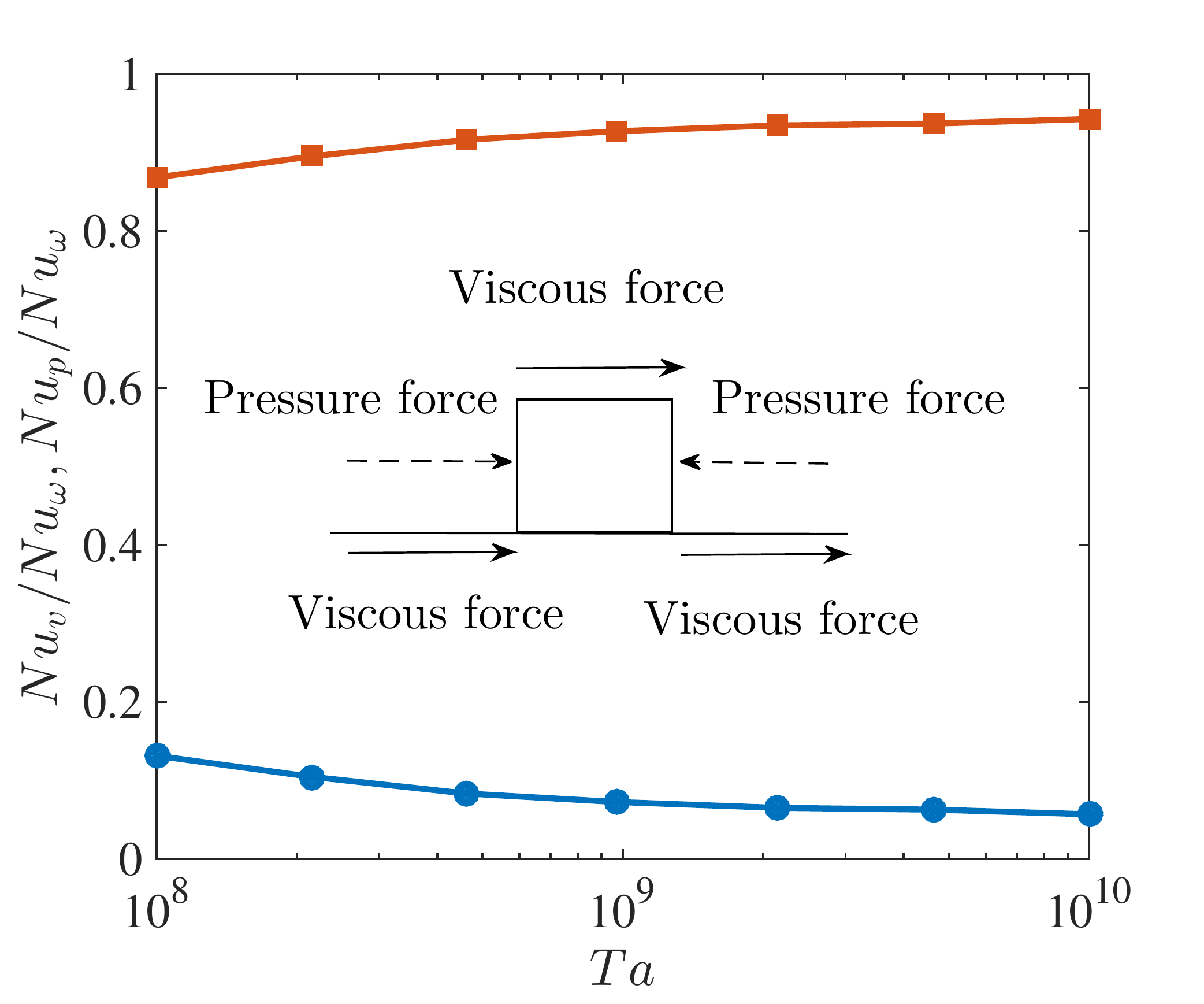}
\caption{Contributions to the total torque originating from pressure $Nu_p$ (upper curve, squares) and viscous stresses $Nu_v$ (lower curve, dots) for rough boundaries. As seen, the torque mainly originates from pressure forces whereas viscous stresses contribute only for a small faction. The schematics inside the figure show how viscous forces and pressure contribute separately to the total torque.}
\label{fig2}
\end{figure}

\begin{figure}

\begin{center}

    \includegraphics[width=2.8in]{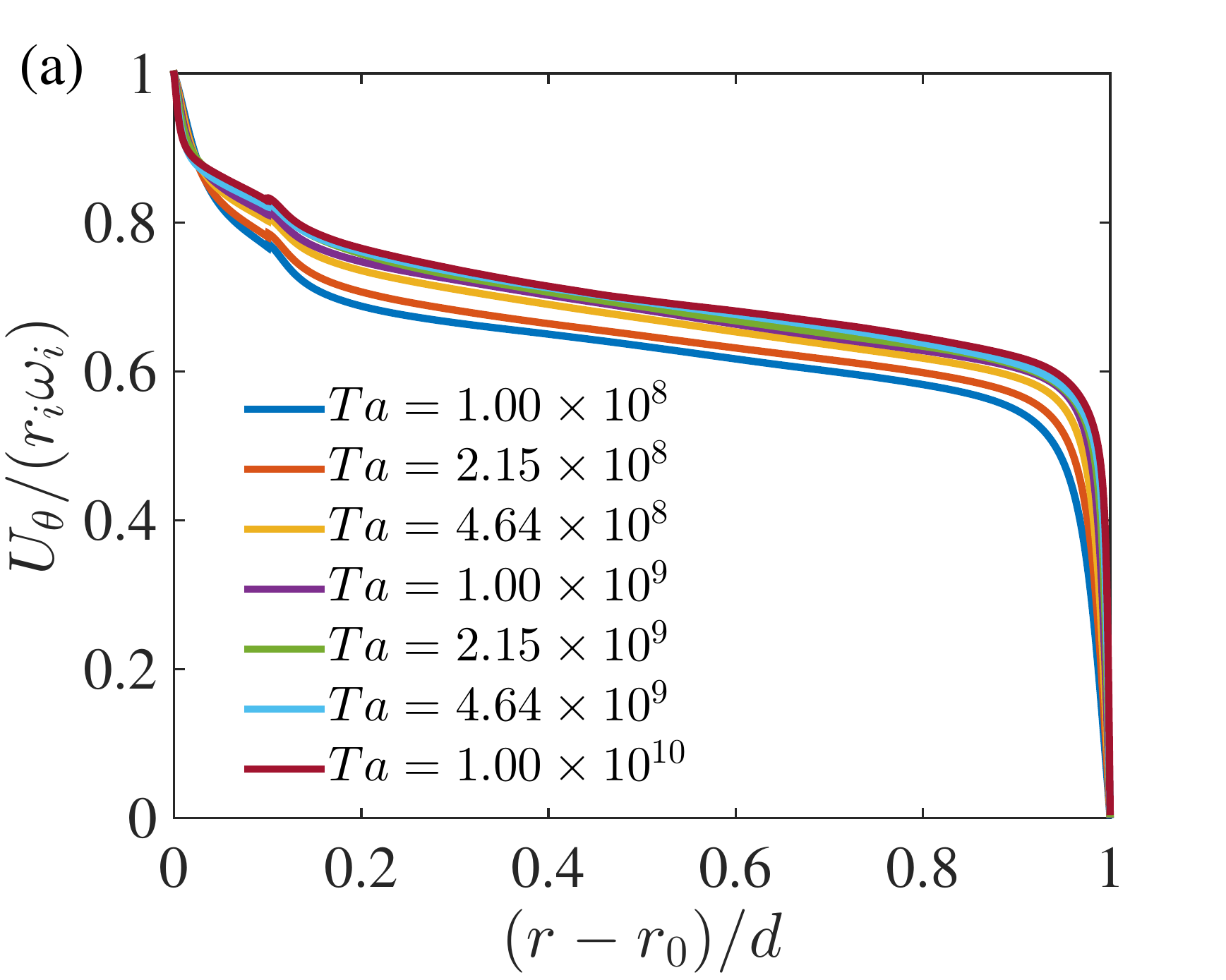}
\end{center}

  \begin{minipage}[c]{0.49\textwidth}
    \includegraphics[width=2.8in]{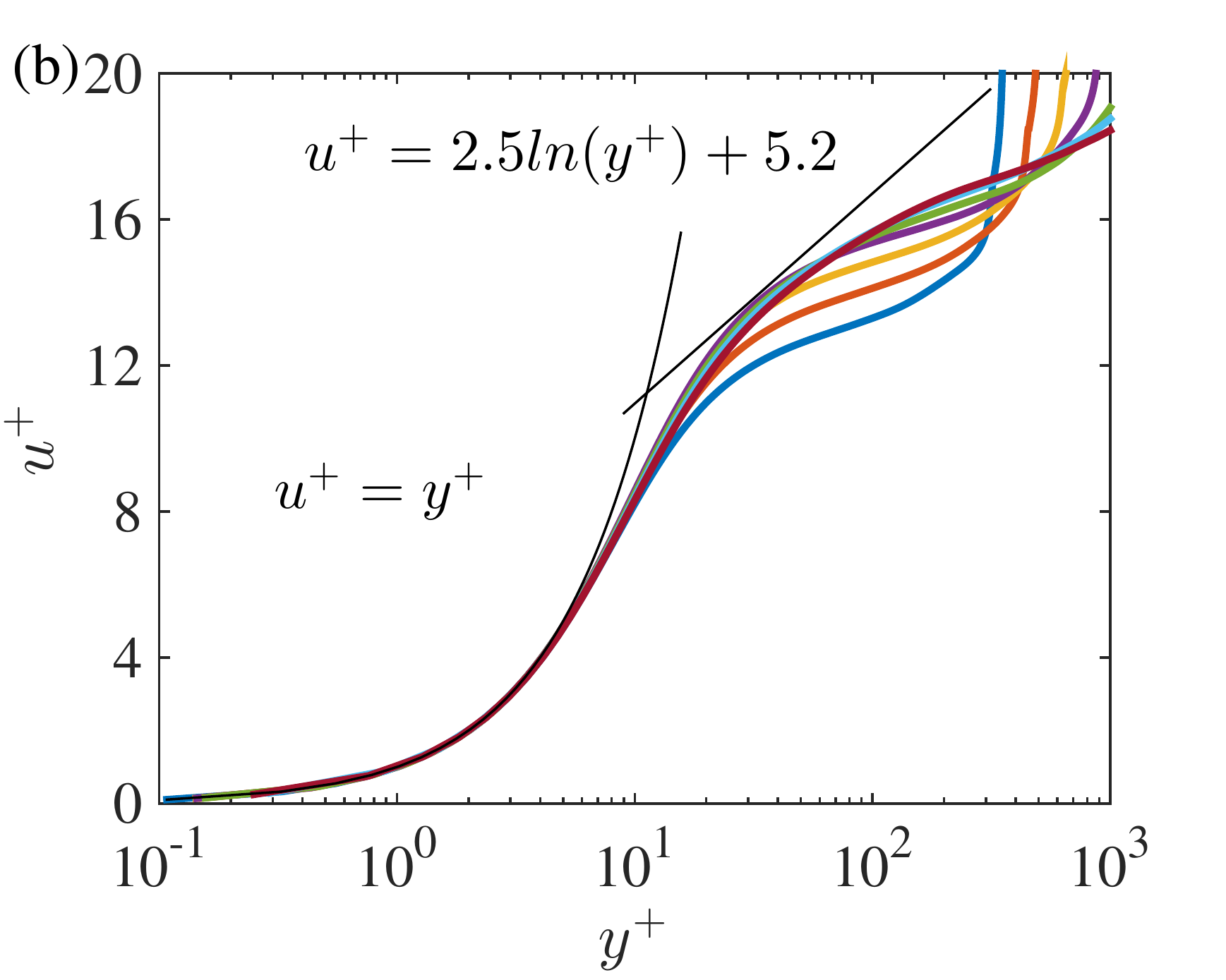}
    
  \end{minipage}
  \begin{minipage}[c]{0.49\textwidth}
    \includegraphics[width=2.8in]{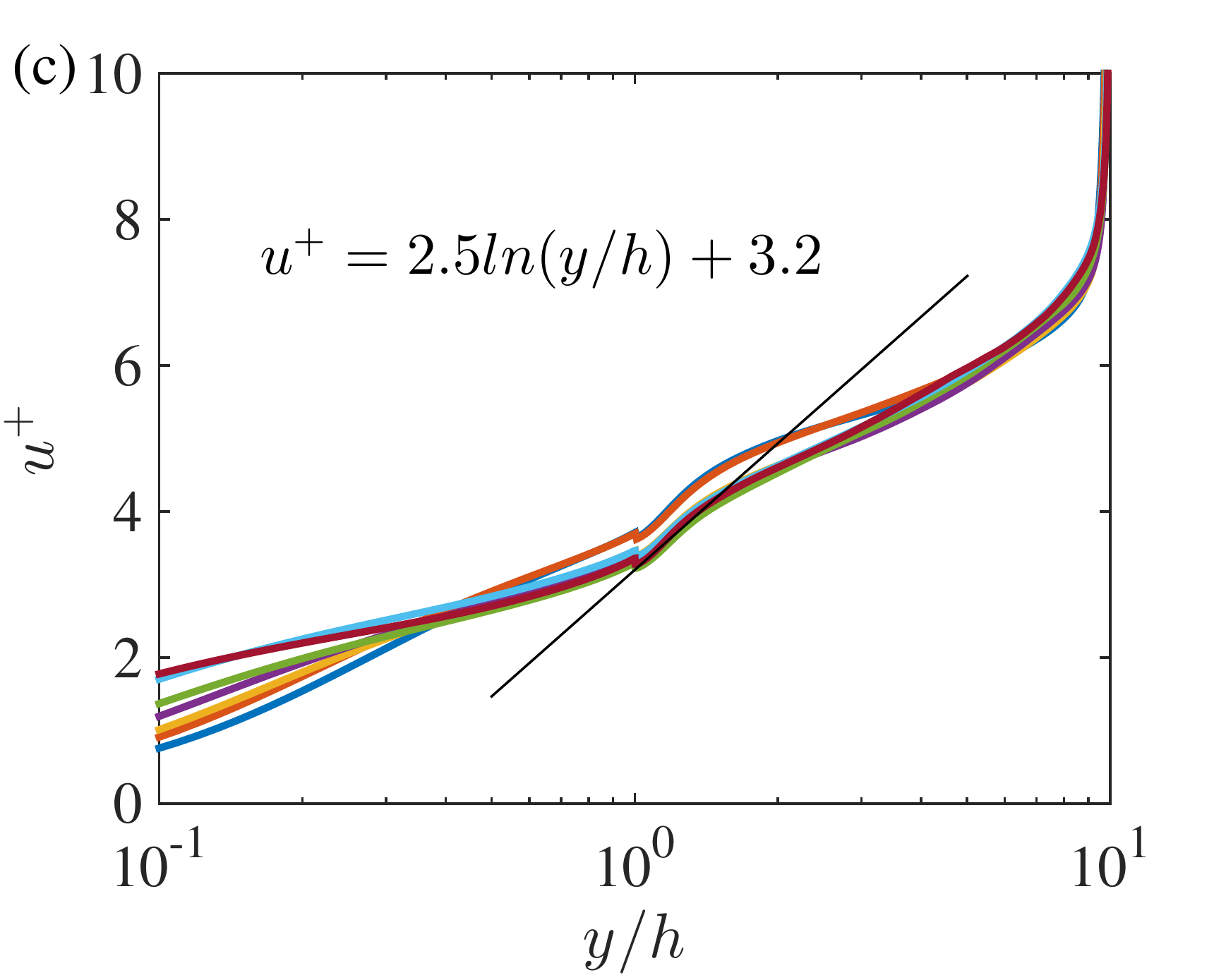}
    
  \end{minipage}
\caption{(a) Averaged azimuthal velocity profiles for varying $Ta$. The shear rate is much smaller at the rough wall compared to the smooth one. The velocity in the middle of the gap is biased towards the rough wall velocity. The larger $Ta$, the more the velocity profile is biased in the bulk towards the inner cylinder velocity. (b) Azimuthal velocity profile non-dimensionalized by wall units at the smooth outer boundary. There are lines which show the law of the wall $u^+=\kappa^{-1} \mathrm{ln} y^++B$, with values of $\kappa=0.4$ and $B=5.2$, and viscous sublayer $u^+=y^+$. (c) Non-dimensionalized azimuthal velocity profile at the rough inner boundary. Note that now the profiles are scaled with the outer variable $y/h$, and not $y^+$. The solid line shows that the law of the wall with roughness in the fully rough regime is $u^+=\kappa^{-1} \mathrm{ln}(y/h)+B$, with the same $\kappa=0.4$ and the smaller $B=3.2$.}
\label{fig3}
\end{figure}

With respect to the global transport properties, the most distinctive modification produced by the roughness is that the torque (at the rough boundary) consists of two contributions, namely viscous stresses and pressure forcing while for the smooth case it is solely the former one. When the roughness scale $h$ is large compared to the viscous length scale $\delta_\nu$, the local Reynolds number of the flow over the rough elements is large, i.e. $u_\tau h/ \nu=h/\delta_\nu \gg 1$. For our lowest $Ta$, this value is already around 50, while for the highest, it is about 400. Almost all of our simulations are thus in the fully rough regime where $h/\delta_\nu>70$ \citep{schlichting_book}. The transfer of momentum from the wall to the fluid is accomplished by the torque of the rough elements, which at high Reynolds number is dominated by the pressure forces, rather than by viscous stresses. This statement is made quantitative in figure \ref{fig2}, showing the decomposition of the total torque into viscous and pressure contributions at the rough boundary. The viscous forces part is defined as 
\begin{eqnarray}
Nu_{\nu}=\int \frac{\tau_\nu r}{\tau_{pa}}  dS,
\end{eqnarray} where $\tau_\nu$ is the viscous shear stress and $r$ the radius, while the pressure part is defined as 
\begin{eqnarray}
Nu_p=\int \frac{P r}{\tau_{pa}}  dS,
\end{eqnarray} being $P$ is the pressure. An illustration of how pressure and viscous forces contribute to the torque separately is shown in figure \ref{fig2}.
In fact, the viscous stresses contribute only for a small fraction to $Nu_\omega$, owing to the recirculations around the rough element, while the pressure forces on the radial faces of the rough elements contribute nearly $95\%$ of the total.

\subsection{Velocity profiles}

The implication of the reduced viscous stresses contribution to the total torque is that the shear rate of the azimuthal velocity at the rough wall becomes smaller compared to the smooth case. As a result, the azimuthal velocity at the mid gap should be biased towards the rough wall relative to the smooth case, reflecting the stronger coupling of the rough wall to the bulk. This is evident from figure \ref{fig3}a, which clearly shows that for increasing $Ta$, the azimuthal velocity in the bulk gets closer to the rough wall velocity. The biased velocity phenomenon was also hypothesized by \cite{berg2003} by using a circuit analogy.

Next, we compare the differences between azimuthal velocity profiles $U_\theta$ normalized in the form $u^+=U_\theta /u_\tau$ versus the wall distance $y$ expressed in terms of $y^+=y/\delta_\nu$ at smooth and rough boundaries, see figure
\ref{fig3}b, c. The outer cylinder boundary layer is that of a smooth wall for which it is well known that there is a viscous sublayer ($u^+=y^+$) for $y^+ < 5$ and a logarithmic profile
 \begin{eqnarray}\label{EQ6}
u^+=\frac{1}{\kappa} \mathrm{ln} y^++B,
\end{eqnarray}
 for $y^+ > 30$. In the above equation, following \cite{huisman2013}, we use $\kappa=0.4$, $B=5.2$ and figure \ref{fig3}b shows convergence for increasing $Ta$ although substantial deviation is evident which originates from the boundary curvature and the Taylor rolls \citep{huisman2013,grossmann2014,grossmann2016,ostilla2016}. 

\begin{figure}

 \centering

     \includegraphics[width=2.8in]{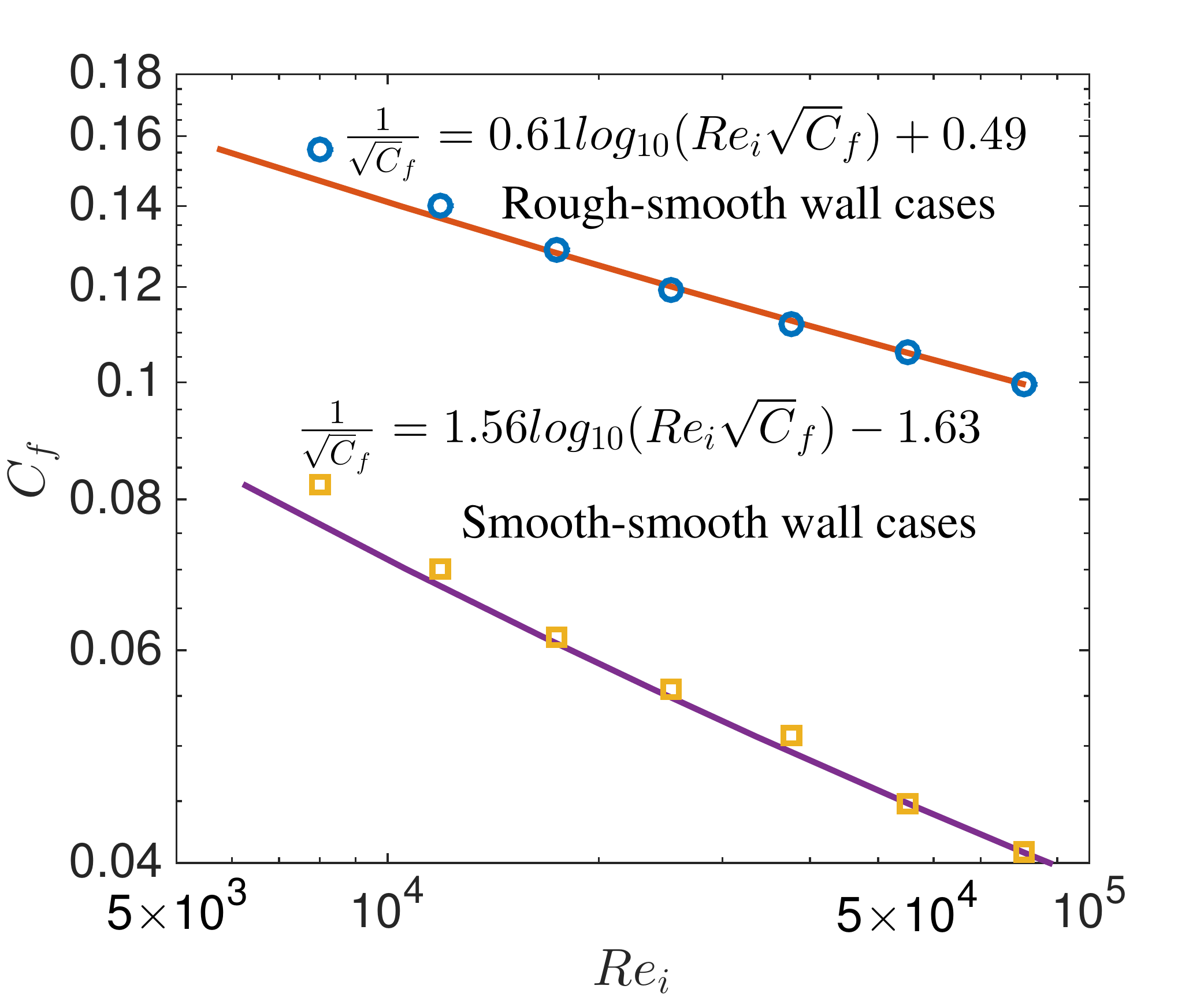}

\caption{Friction factor $C_f=2\pi Nu_\omega J^\omega_0 \nu^{-2}/Re_i^2$ as a function of the inner cylinder Reynolds number $Re_i$, where $J^\omega_0=2\nu r_o^2 r_i^2 \omega_i/(r_o^2-r_i^2)$ is the angular velocity current at the purely azimuthal laminar state. The circles denote the data extracted from DNS for our current rough-smooth (RS) wall cases, and the squares the data for smooth-smooth (SS) wall cases from \cite{ostilla2014a,ostilla2014b}, respectively. The lines show the best fits of the Prandtl-von K\'arm\'an law $1/\sqrt C_f=a \mathrm{log}_{10}(Re_i\sqrt C_f)+b$, with $a=0.61$, $b=0.49$ for RS cases and $a=1.56$, $b=-1.63$ for SS cases.}
\label{6}
\end{figure}

At the inner cylinder rough boundary, the law of the wall is extended to incorporate roughness. Because of the pressure forces dominance, $\delta_\nu$ is no longer the relevant parameter. Instead, the roughness scale $h$ should be used to normalize the wall distance \citep{pope_book}. With this change, the log-law with roughness in the fully rough regime \citep{pope_book,nikuradse1933} becomes 
\begin{eqnarray}\label{EQ7}
u^+=\frac{1}{\kappa} \mathrm{ln} \left( \frac{y}{h} \right)+B,
\end{eqnarray}
to be compared with (\ref{EQ6}). In figure \ref{fig3}c, above the roughness height, there is a region ($0.1<y/h<0.16$) where the velocity profiles collapse very well and show the logarithmic behaviour Eq. (\ref{EQ7}), with the von K\'arm\'an prefactor. The von K\'arm\'an constant $\kappa$ stays the same while the fit for $B$ gives $B=3.2$, exactly the same as found in rough channel flow \citep{ikeda2007}.

\subsection{Skin friction factor}
An alternative way to describe the global transport properties is through the friction factor $C_f$
as a function of the inner cylinder Reynolds number $Re_i$, as shown in figure \ref{6}. The Prandtl-von K\'arm\'an skin friction law \citep{schlichting_book,pope_book,lewis1999,berg2003}, which is an extension of the log law of the wall, is used to fit our data. The implicit form is 
\begin{eqnarray}\label{EQ8}
\frac{1}{\sqrt C_f}=a \mathrm{log}_{10}(Re_i\sqrt C_f)+b.
\end{eqnarray}
For the smooth-smooth (SS) wall case, we get $a=1.56$ and $b=-1.63$. This differs from the value of $a$ from turbulent channel ($a=1.91$, see \cite{zanoun2009}) or pipe flows ($a=1.92$, see \cite{mckeon2005}), due to the effects of curvature and large scale Taylor rolls, which make the Karman constant slightly larger in TC flow \citep{ostilla2014a,huisman2013}. The best fit for the current rough-smooth (RS) wall case results in $a=0.61$ and $b=0.49$. Compared to $a=1.56$ in TC turbulence with the smooth-smooth (SS) wall, here for RS cases $a$ is much smaller. It is important to note that on the one hand, the velocity profile for the RS cases is a combination of smooth-wall Eq. (\ref{EQ6}) and rough-wall Eq. (\ref{EQ7}) situations, on the other hand, the logarithmic term of Eq. (\ref{EQ7}) is independent of $Re_i$. Therefore, the dependence of $C_f$ on $Re_i$ weakens for the RS cases as compared to the SS cases.

 \begin{figure}
\centering\includegraphics[width=3.in]{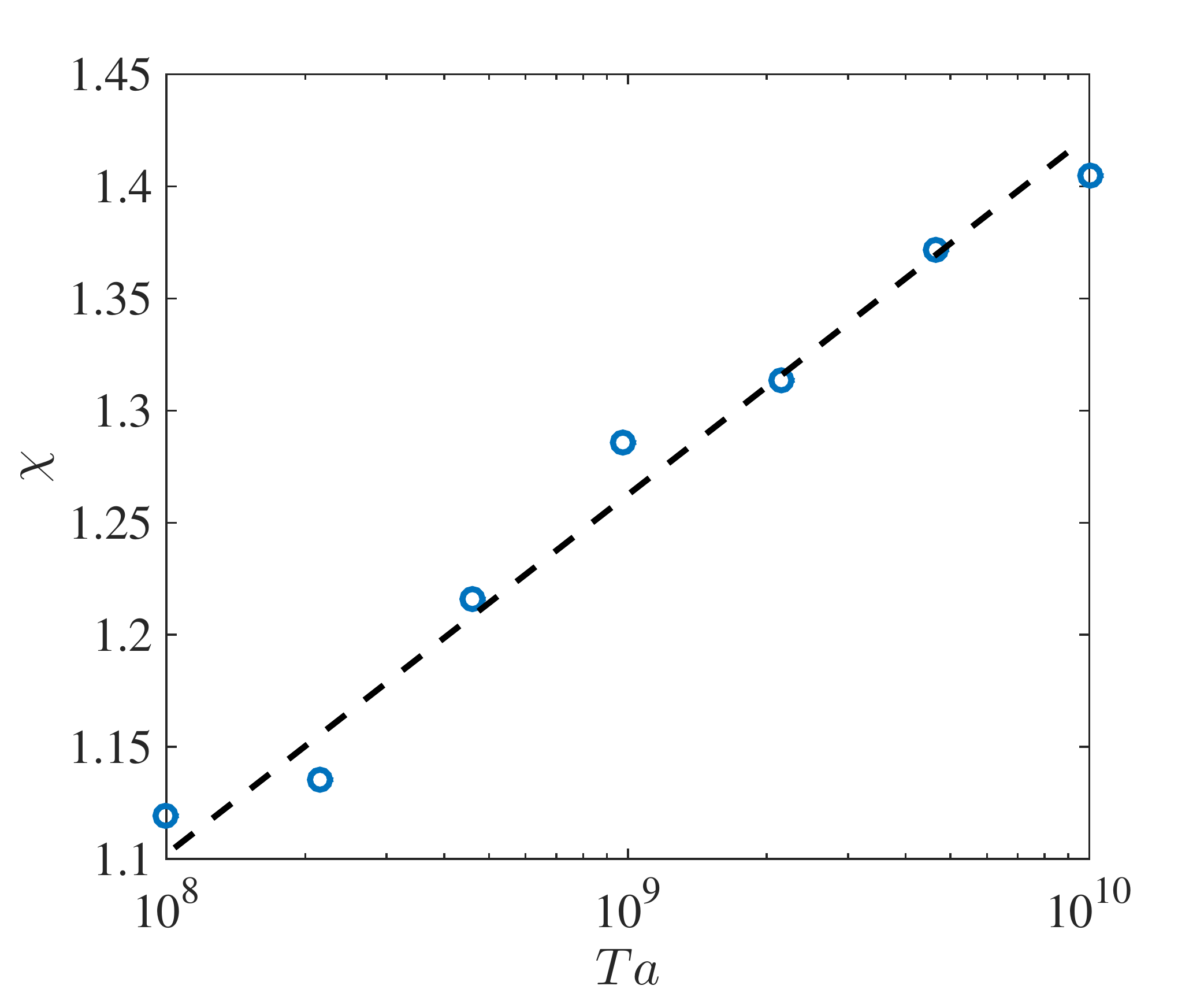}
\caption{Ratio $\chi$ between the angular velocity drop of the smooth wall and the rough wall as a function of $Ta$. The dashed line shows the trend of the asymmetry, which increases with increasing $Ta$.}
\label{fig7}
\end{figure}

\subsection{Asymmetry velocity profiles between smooth and rough walls}
\begin{figure}
 \begin{minipage}[c]{0.5\textwidth}
    \includegraphics[width=2.8 in]{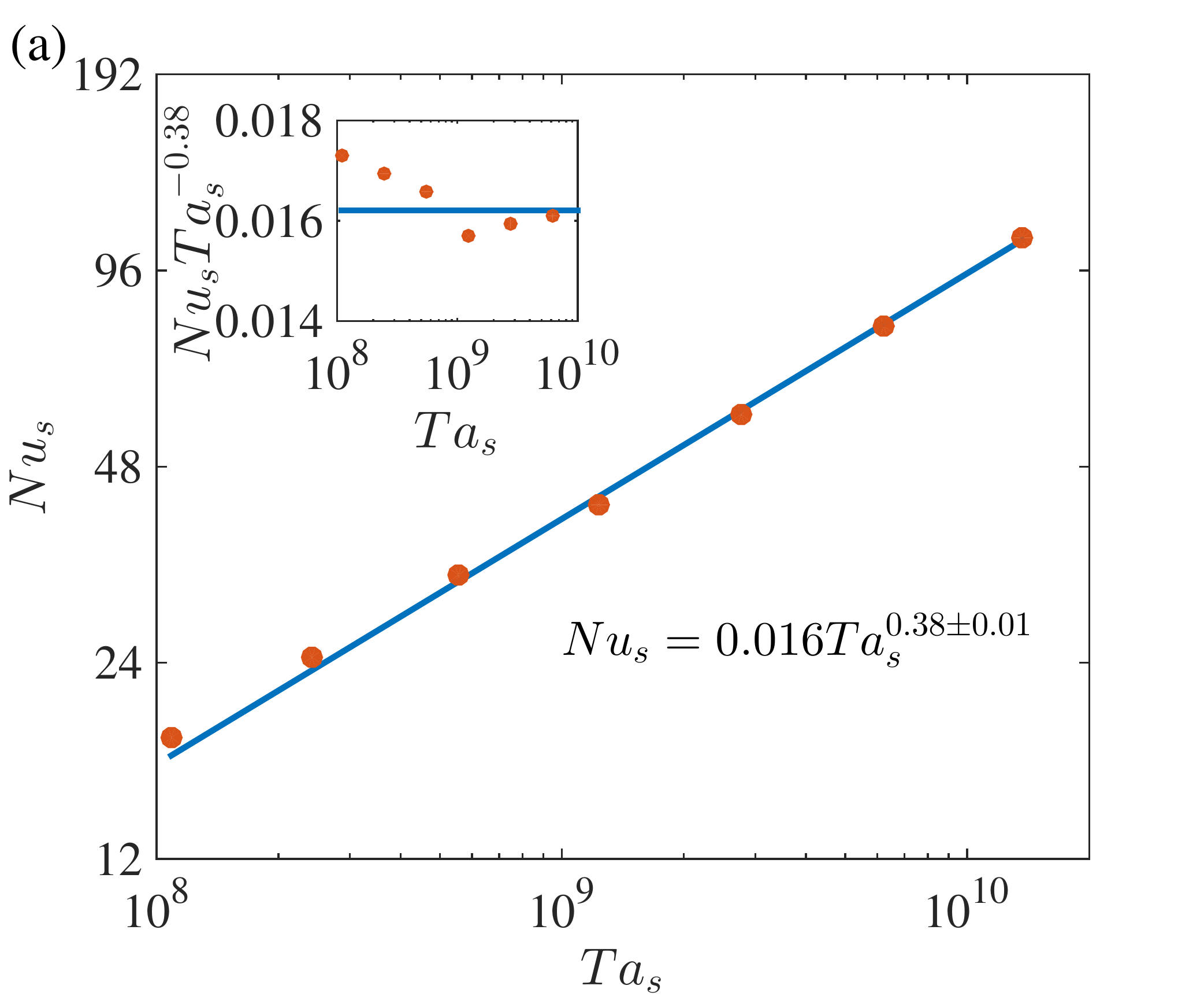}
   
  \end{minipage}%
     \begin{minipage}[c]{0.5\textwidth}
    \includegraphics[width=2.8 in]{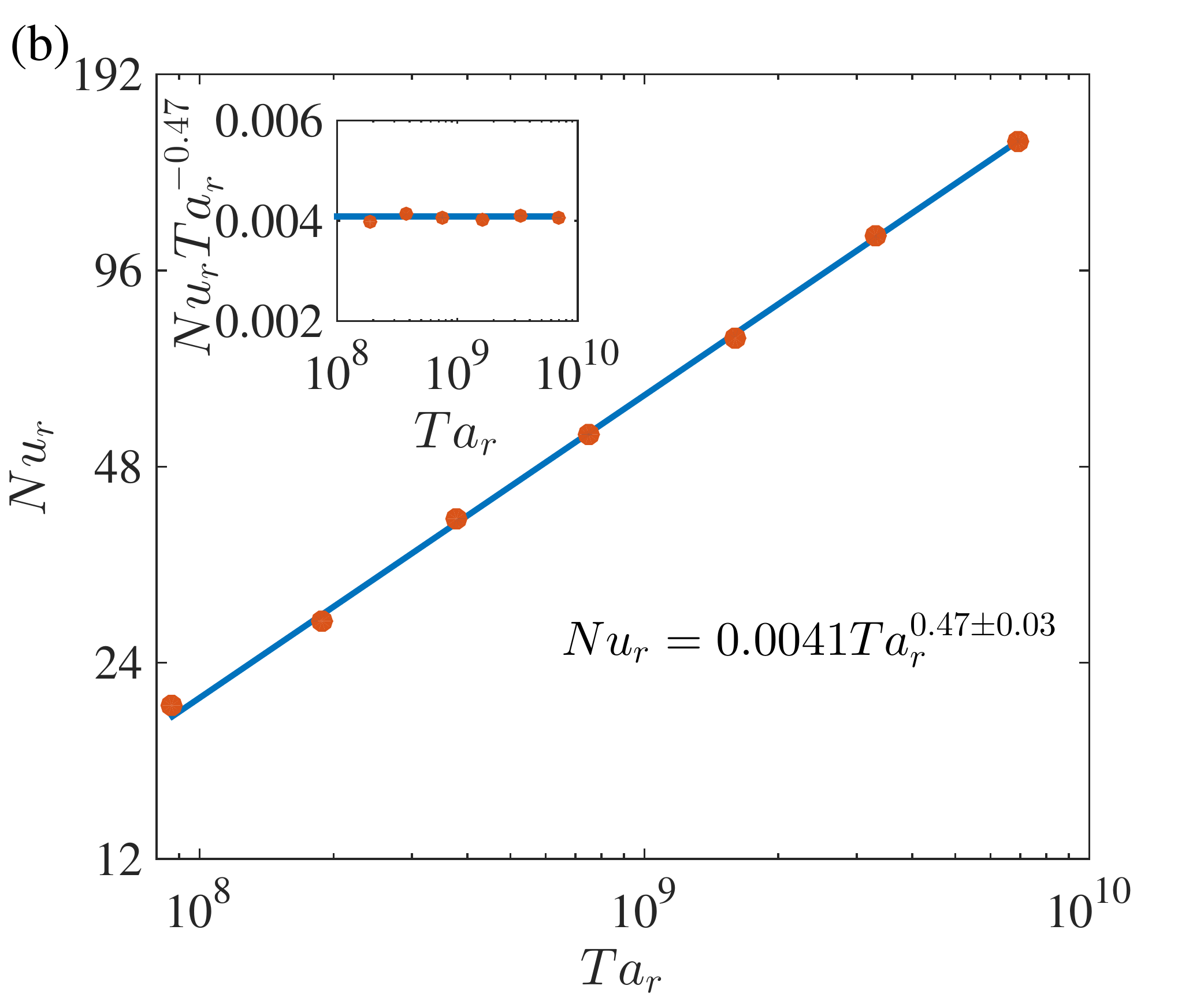}
    
  \end{minipage}
    \caption{(a) Smooth wall Nusselt number $Nu_s$ as a function of smooth wall Taylor number $Ta_s$. (b) Rough wall Nusselt number $Nu_r$ as a function of rough wall Taylor number $Ta_r$. The bullets are the calculated $Nu_s$ and $Nu_r$ from simulations while the straight lines are the best fits $Nu_s=0.016Ta_s^{0.38\pm0.01}$ for the smooth wall and $Nu_r=0.0041Ta_r^{0.47\pm0.03}$ for the rough wall. The insets show the corresponding compensated plots. }
\label{fig6}
\end{figure}

As stated before, we find the velocity profiles biased towards the rough wall. The lager $Ta$ is, the more asymmetric the velocity profile is. To quantify this, a asymmetry parameter $\chi$ is introduced, defined as
\begin{eqnarray}\label{EQ9}
\chi=\frac{\omega_m }{\omega_i-\omega_m}.
\end{eqnarray}
Here, $\omega_m$ is the angular velocity at the middle gap. To repeat symmetry, one would have $\chi=1$. The numerically obtained values of $\chi$ as function of $Ta$ is shown in figure \ref{fig7}. By using one rough wall and one smooth wall, we can compare the transport
properties of different walls within the same 
flow configuration. This has been done in RB flow by \cite{tisserand2011} and \cite{wei2014}. Similarly for TC flow, if assuming that a symmetric cell has two independent smooth walls behaving the same as our outer cylinder, we get the difference of angular velocity across the gap as $2\omega_m$. Thus, the smooth wall Nusselt number is 
\begin{eqnarray}\label{EQ10}
Nu_s=\frac{Nu_\omega \omega_i}{2\omega_m}=\frac{Nu_\omega }{2} \left(\frac{1+\chi}{\chi}\right),
\end{eqnarray}
and the corresponding smooth wall Taylor number $Ta_s$ is 
\begin{eqnarray}\label{EQ11}
Ta_s=\frac{4Ta\omega_m^2}{\omega_i^2}=\frac{4Ta\chi^2}{(1+\chi)^2},
\end{eqnarray}
both restore to the symmetric case for $\chi=1$. For the rough boundary, a rough wall Nusselt number is defined as 
\begin{eqnarray}\label{EQ12}
Nu_r=\frac{Nu_\omega \omega_i}{2(\omega_i-\omega_m)}=\frac{Nu_\omega}{2}(1+\chi),
\end{eqnarray}
with its rough wall Taylor number 
\begin{eqnarray}\label{EQ13}
Ta_r=\frac{4Ta(\omega_i-\omega_m)^2}{\omega_i^2}=\frac{4Ta}{(1+\chi)^2},
\end{eqnarray}
again, both restore to symmetric case for $\chi=1$. It is easy to find that 
\begin{eqnarray}\label{EQ14}
Nu_r=\chi Nu_s,
\end{eqnarray}
and
\begin{eqnarray}\label{EQ14}
Ta_r=\frac{ Ta_s}{\chi^2}.
\end{eqnarray} 
Note that from figure \ref{fig7} a rather linear growth of $\chi$ over two decades of $Ta$ is found, which would surely make the effective scalings between $Nu_s$ vs. $Ta_s$ and $Nu_r$ vs. $Ta_r$ different. As a confirmation, in figure \ref{fig6},  $Nu_s$ vs. $Ta_s$ and $Nu_r$ vs. $Ta_r$ are shown. For the smooth wall, a clear scaling of $Nu_s\propto Ta_s^{0.38\pm0.01}$ is revealed, which is in excellent agreement with the case where both walls are smooth and corresponds to the effective scaling of ultimate regime with logarithmic correction \citep{grossmann2011,vangils2011,huisman2012}. This indicates that the rough wall can not affect the scaling of the smooth wall, even in the ultimate regime. In contrast, the rough wall results show a scaling of $Nu_r\propto Ta_r^{0.47\pm0.03}$, which is very close to the $1/2$ scaling proposed by \cite{kraichnan1962}, corresponding to the pure ultimate regime without logarithmic correction. The competition between the smooth and rough walls eventually determines the global scaling exponent to be 0.42.

\begin{figure}
\centering\includegraphics[width=3.in]{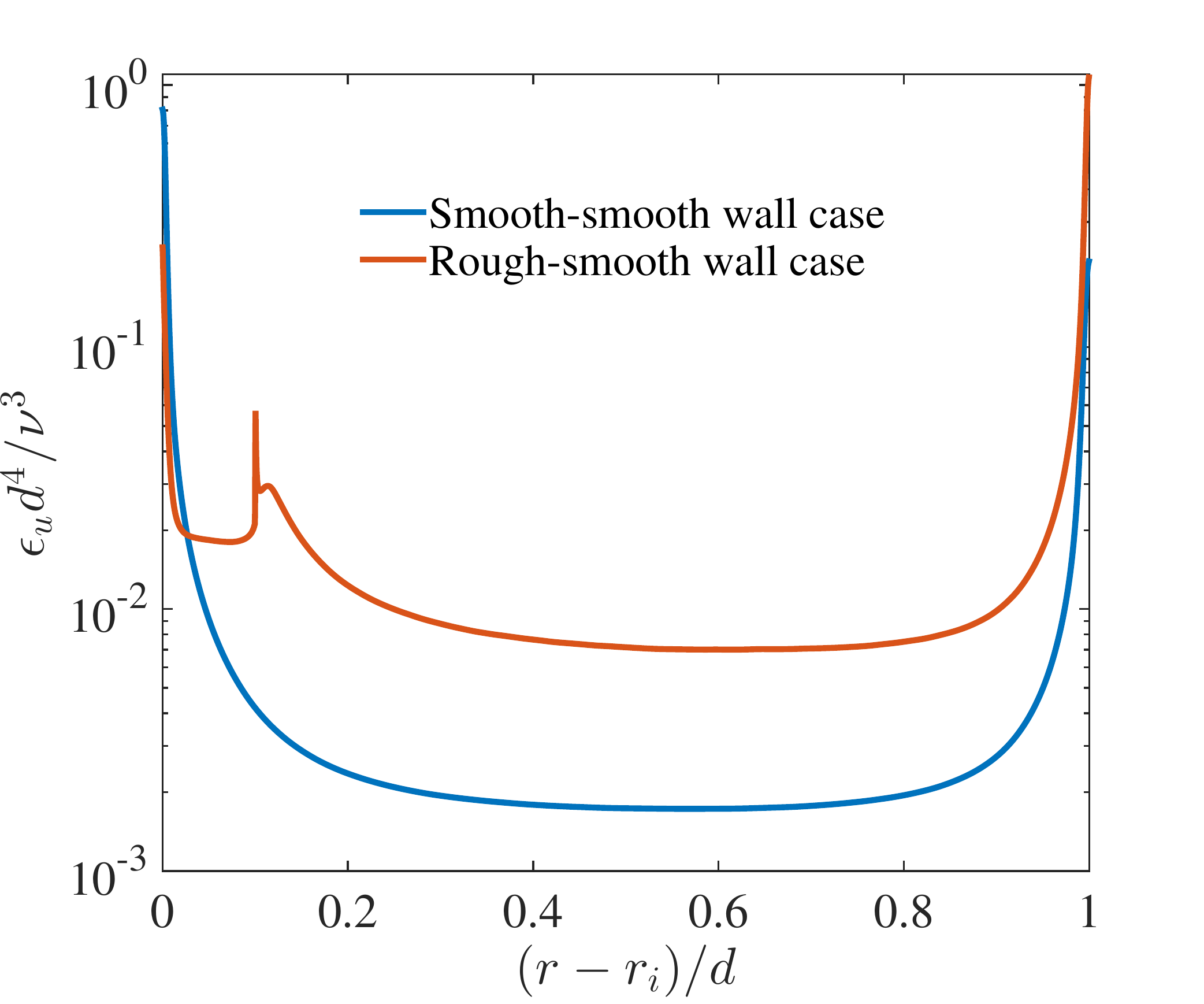}
\caption{Dimensionless averaged energy dissipation rate $\epsilon_ud^4/\nu^3$ distribution along the radial direction at the current RS case in comparison with the SS case at $Ta=4.6\times10^9$. The dissipation is greatly enhanced at the bulk while decreased close to the rough boundary in the RS case as compared to the SS case.}
\label{fig5}
\end{figure}

\begin{figure}


    \hspace*{-0.25in} \includegraphics[width=5.8in]{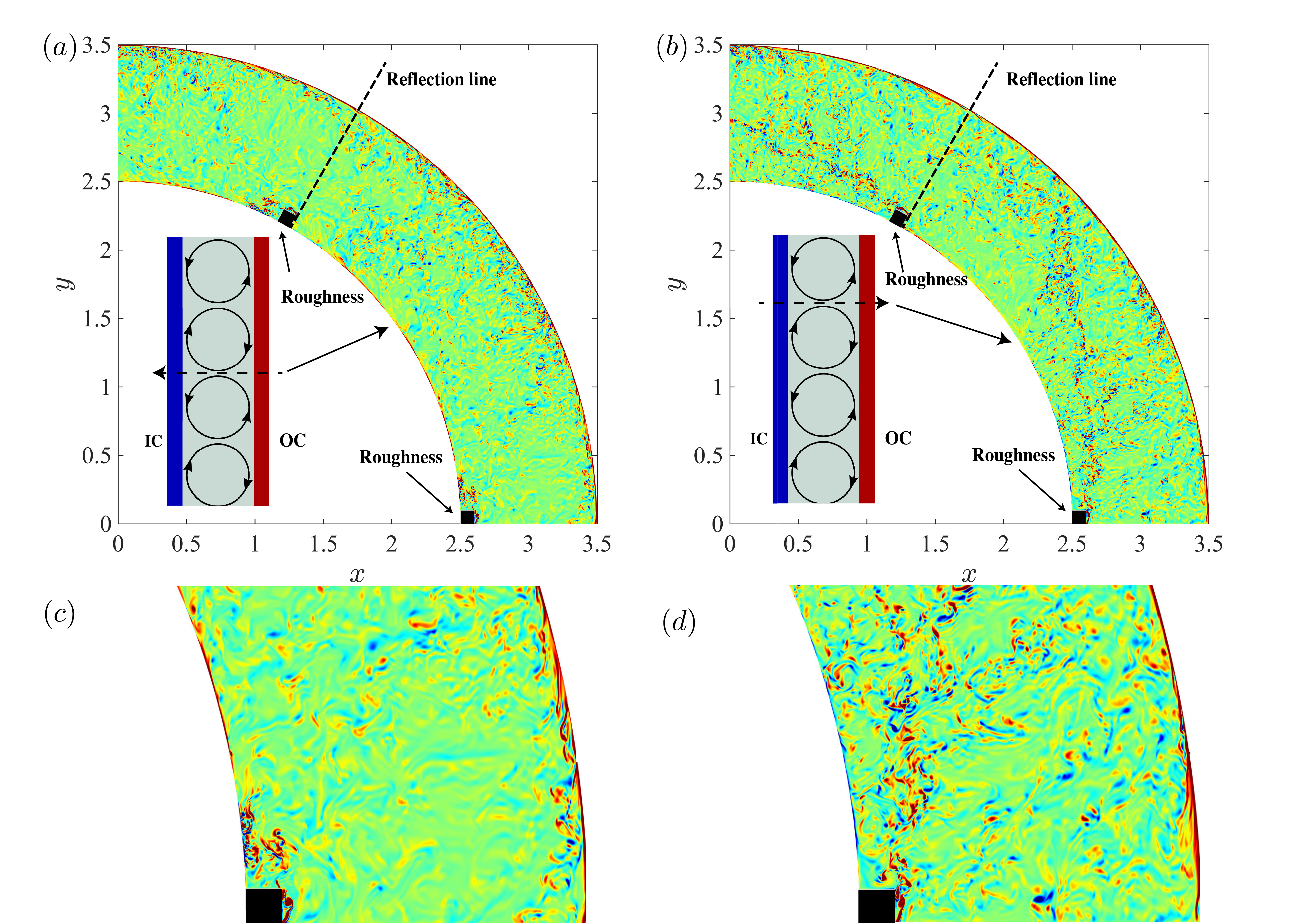}

\caption{Instantaneous axial vorticity field for $Ta=4.64\times10^9$ at two heights for the current inner rough wall case. The insets show schematic sketches of Taylor rolls in the meridional plane. The dashed lines denote where the cross section contour plots are intercepted. The arrows on the dashed lines show the directions of the radial velocities. (a) At the height where Taylor roll is circulating from the outer to the inner cylinder. (b) At the height where Taylor roll is moving from the inner to the outer cylinder. Vortex-shedding is found at the top of the rough element due to flow separation either to the back of the rough element (a) or to the bulk (b). Note that the reflection lines are used to show the regions beyond the computation domain.  Abbreviations: IC (inner cylinder) and OC (outer cylinder). (c) Enlargement
the near wall region for (a). (d) Enlargement
the near wall region for (b).}
\label{fig4}
\end{figure}

\subsection{Energy dissipation rate}

Like RB flow, in TC flow an exact relation between the mean kinetic energy dissipation $\epsilon_{u,m}$ and the driving force $Ta$ holds, expressed as \citep*{eckhardt2007b}
\begin{eqnarray}\label{EQ14}
\epsilon_{u,m}=\nu^3 d^{-4}Ta(Nu_\omega-1)\left (\frac{\sqrt \eta}{(1+\eta)/2} \right)^8 .
\end{eqnarray} 
Here we show how the inner wall roughness alters the kinetic energy dissipate rate distribution along the radial direction and thus increases the scaling exponent. Again due to the pressure dominance, the dissipation at the rough wall is decreased significantly as compared to the smooth wall case where the shear rate at the wall is extremely large. These effects redistribute the energy dissipation rate along the radius. Figure \ref{fig5} details the comparison of the energy dissipation rate distribution along the radius in our current RS and SS cases at $Ta=4.64\times10^9$. Applying the Grossmann-Lohse theory \citep{grossmann2000,grossmann2001} here, we separate the energy dissipation rate as boundary layer and bulk contributions. The more the bulk is dominant, the larger the scaling exponent is \citep{grossmann2004}. The boundary layer thickness is estimated by $\lambda/d=1.058/(2Nu_\omega)$ for the current radius ratio \citep{braukmann2013}. For $Ta=4.64\times10^9$, it is found that at the case with both smooth walls, the two boundary layers contribute 65\% of the total dissipation while the bulk contributes 35\%. With the rough inner wall, the dissipation generated below the height of the rough element only contributes 11\% to the total dissipation and the outer smooth boundary generates 26\%. Thus the bulk contributes 63\%. Therefore, in contrast to the smooth case, TC turbulence with inner rough wall is more bulk dominant and hence the torque scaling exponent is increased.

\subsection{Vorticity field}

To shed further light on how roughness pushes the dissipation into the bulk, in figure \ref{fig4}, we show the instantaneous axial vorticity field for $Ta=4.64\times10^9$ at two sections: one for the Taylor rolls producing a current from inner to the outer cylinder and the other in the opposite direction. It is found that due to the smaller shear rate, very few vortices are generated at the surfaces behind the rough elements compared to the smooth wall. Instead, flow separations at the top of the rough elements cause vortex-shedding from it. These shed vortices are then transported by Taylor rolls either to the outer cylinder, which enhance dissipation in the bulk region, or to the cavities between the rough elements at the inner cylinder, which contribute to the dissipation in the near wall region as well. That explains why although the shear rate is decreased significantly, the dissipation below the height of the rough element still contributes 11\% of the total.

\section{Conclusion}

In summary, we have performed DNSs of TC turbulence with inner rough wall. It is found that the wind Reynolds number scales as $Re_w \propto Ta^{0.50\pm0.01}$, in accordance with the prediction by \cite{grossmann2011} for smooth boundaries. The angular velocity flux is found to have an effective scaling of $Nu_\omega \propto Ta^{0.42\pm0.01}$, which is a notably higher effective exponent than the ultimate regime value 0.38 seen for smooth walls. Furthermore, the dominant torque at the rough boundary stems from the pressure forces on the side faces of the rough element, rather than from viscous stresses. As a consequence, the law of the wall at the rough boundary depends on the roughness height, rather than the viscous length scale. By separating the torque into the smooth and rough walls contributions, we find that the smooth wall torque scaling follows $Nu_s \propto Ta_s^{0.38\pm0.01}$, the same as with both smooth walls. In contrast, the rough wall scaling follows $Nu_r \propto Ta_r^{0.47\pm0.03}$, close to the pure ultimate scaling $Nu_\omega \propto Ta^{1/2}$. Lastly, the wall value of the energy dissipation rate decreases due to the smaller shear rate at the rough boundary while dissipation is redistributed through vortex-shedding into the bulk. The system becomes bulk dominant, thus reducing the logarithmic correction and increasing the effective torque scaling exponent.

Note that the scaling exponent increase is only possible as the pressure contribution dominates the torques. Therefore it is interesting to try different spacings among the rough elements for either pressure or viscous force dominance. By doing so we may be able to control the torque scaling exponent in TC turbulence.  
Finally, a subsequent study of TC turbulence with both rough walls now seems mandatory in order to show whether the pure ultimate scaling $Nu_\omega \propto Ta^{1/2}$ without logarithmic correction is achievable or not. 

\bigskip
We thank R. Verschoof for valuable discussions. This work is supported by FOM and MCEC, both funded by NWO. We thank the Dutch Supercomputing Consortium
SurfSARA, the Italian supercomputer FERMI-CINECA through the PRACE Project
No. 2015133124 and the ARCHER UK National Supercomputing Service through the DECI Project 13DECI0246 for the allocation of computing time.

\bibliographystyle{jfm}
\bibliography{jfm-instructions}

\begin{thebibliography}{42}
\expandafter\ifx\csname natexlab\endcsname\relax\def\natexlab#1{#1}\fi
\def\au#1{#1} \def\ed#1{#1} \def\yr#1{#1}\def\at#1{#1}\def\jt#1{\textit{#1}}
  \def\bt#1{#1}\def\bvol#1{\textbf{#1}} \def\vol#1{#1} \def\pg#1{#1}
  \def\publ#1{#1}\def\arxiv#1{#1}\def\org#1{#1}\def\st#1{\textit{#1}}

\bibitem[Ahlers {\em et~al.\/}(2009)Ahlers, Grossmann \& Lohse]{ahlers2009}
{\sc \au{Ahlers, G.}, \au{Grossmann, S.} \& \au{Lohse, D.}} \yr{2009}  \at{Heat
  transfer and large scale dynamics in turbulent {Rayleigh-B\'{e}nard}
  convection}.  \jt{Rev. Mod. Phys.}  \bvol{81},  \pg{503--537}.

\bibitem[van~den Berg {\em et~al.\/}(2003)van~den Berg, Doering, Lohse \&
  Lathrop]{berg2003}
{\sc \au{van~den Berg, T.}, \au{Doering, C.}, \au{Lohse, D.} \& \au{Lathrop,
  D.}} \yr{2003}  \at{Smooth and rough boundaries in turbulent {Taylor-Couette}
  flow}.  \jt{Phys. Rev. E}  \bvol{68},  \pg{036307}.

\bibitem[Brauckmann \& Eckhardt(2013)]{braukmann2013}
{\sc \au{Brauckmann, H.~J.} \& \au{Eckhardt, B.}} \yr{2013}  \at{Direct
  numerical simulations of local and global torque in {Taylor-Couette} flow up
  to {$Re = 30~000$}}.  \jt{J. Fluid Mech.}  \bvol{718},  \pg{398--427}.

\bibitem[Busse(2012)]{busse2012}
{\sc \au{Busse, F.}} \yr{2012}  \at{Viewpoint: The twins of turbulence
  research}.  \jt{Physics}  \bvol{5}.

\bibitem[Cadot {\em et~al.\/}(1997)Cadot, Couder, Daerr, Douady \&
  Tsinober]{cadot1997}
{\sc \au{Cadot, O.}, \au{Couder, Y.}, \au{Daerr, A.}, \au{Douady, S.} \&
  \au{Tsinober, A.}} \yr{1997}  \at{Energy injection in closed turbulent flows:
  Stirring through boundary layers versus inertial stirring}.  \jt{Phys. Rev.
  E}  \bvol{56},  \pg{427--433}.

\bibitem[Ciliberto \& Laroche(1999)]{ciliberto1999}
{\sc \au{Ciliberto, S.} \& \au{Laroche, C.}} \yr{1999}  \at{Random roughness of
  boundary increases the turbulent scaling exponents}.  \jt{Phys. Rev. Lett.}
  \bvol{82},  \pg{3998--4001}.

\bibitem[Du \& Tong(2000)]{du2000}
{\sc \au{Du, Y.-B.} \& \au{Tong, P.}} \yr{2000}  \at{Turbulent thermal
  convection in a cell with ordered rough boundaries}.  \jt{J. Fluid Mech.}
  \bvol{407},  \pg{57--84}.

\bibitem[Eckhardt {\em et~al.\/}(2007{\natexlab{{\em a\/}}})Eckhardt, Grossmann
  \& Lohse]{eckhardt2007a}
{\sc \au{Eckhardt, B.}, \au{Grossmann, S.} \& \au{Lohse, D.}}
  \yr{2007{\natexlab{{\em a\/}}}}  \at{Fluxes and energy dissipation in thermal
  convection and shear flows}.  \jt{Europhys. Lett.}  \bvol{78},  \pg{24001}.

\bibitem[Eckhardt {\em et~al.\/}(2007{\natexlab{{\em b\/}}})Eckhardt, Grossmann
  \& Lohse]{eckhardt2007b}
{\sc \au{Eckhardt, B.}, \au{Grossmann, S.} \& \au{Lohse, D.}}
  \yr{2007{\natexlab{{\em b\/}}}}  \at{Torque scaling in turbulent
  {Taylor-Couette} flow between independently rotating cylinders}.  \jt{J.
  Fluid Mech.}  \bvol{581},  \pg{221--250}.

\bibitem[Fadlun {\em et~al.\/}(2000)Fadlun, Verzicco, Orlandi \&
  Mohd-Yusof]{fadlun2000}
{\sc \au{Fadlun, E.~A.}, \au{Verzicco, R.}, \au{Orlandi, P.} \& \au{Mohd-Yusof,
  J.}} \yr{2000}  \at{Combined immersed-boundary finite-difference methods for
  three-dimensional complex flow simulations}.  \jt{J. Comput. Phys.}
  \bvol{161},  \pg{35--60}.

\bibitem[van Gils {\em et~al.\/}(2011)van Gils, Huisman, Bruggert, Sun \&
  Lohse]{vangils2011}
{\sc \au{van Gils, D. P.~M.}, \au{Huisman, S.~G.}, \au{Bruggert, G.-W.},
  \au{Sun, C.} \& \au{Lohse, D.}} \yr{2011}  \at{Torque scaling in turbulent
  {Taylor-Couette} flow with co- and counterrotating cylinders}.  \jt{Phys.
  Rev. Lett.}  \bvol{106},  \pg{024502}.

\bibitem[Grossmann \& Lohse(2000)]{grossmann2000}
{\sc \au{Grossmann, S.} \& \au{Lohse, D.}} \yr{2000}  \at{Scaling in thermal
  convection: a unifying theory}.  \jt{J. Fluid Mech.}  \bvol{407},
  \pg{27--56}.

\bibitem[Grossmann \& Lohse(2001)]{grossmann2001}
{\sc \au{Grossmann, S.} \& \au{Lohse, D.}} \yr{2001}  \at{Thermal convection
  for large {Prandtl} numbers}.  \jt{Phys. Rev. Lett.}  \bvol{86},
  \pg{3316--3319}.

\bibitem[Grossmann \& Lohse(2004)]{grossmann2004}
{\sc \au{Grossmann, S.} \& \au{Lohse, D.}} \yr{2004}  \at{Fluctuations in
  turbulent {Rayleigh-B\'enard} convection: The role of plumes}.  \jt{Phys.
  Fluids}  \bvol{16}~(12),  \pg{4462}.

\bibitem[Grossmann \& Lohse(2011)]{grossmann2011}
{\sc \au{Grossmann, S.} \& \au{Lohse, D.}} \yr{2011}  \at{Multiple scaling in
  the ultimate regime of thermal convection}.  \jt{Phys. Fluids}  \bvol{23},
  \pg{045108}.

\bibitem[Grossmann {\em et~al.\/}(2014)Grossmann, Lohse \& Sun]{grossmann2014}
{\sc \au{Grossmann, S.}, \au{Lohse, D.} \& \au{Sun, C.}} \yr{2014}
  \at{Velocity profiles in strongly turbulent {Taylor-Couette} flow}.
  \jt{Phys. Fluids.}  \bvol{26}~(2),  \pg{025114}.

\bibitem[Grossmann {\em et~al.\/}(2016)Grossmann, Lohse \& Sun]{grossmann2016}
{\sc \au{Grossmann, S.}, \au{Lohse, D.} \& \au{Sun, C.}} \yr{2016}
  \at{{High-Reynolds} number {Taylor-Couette} turbulence}.  \jt{Annu. Rev.
  Fluid Mech.}  \bvol{48},  \pg{53--80}.

\bibitem[He {\em et~al.\/}(2012{\natexlab{{\em a\/}}})He, Funfschilling,
  Bodenschatz \& Ahlers]{he2012a}
{\sc \au{He, X.}, \au{Funfschilling, D.}, \au{Bodenschatz, E.} \& \au{Ahlers,
  G.}} \yr{2012{\natexlab{{\em a\/}}}}  \at{Heat transport by turbulent
  {Rayleigh-B\'{e}nard} convection for ${Pr} \simeq 0.8$ and $4 \times 10^{11}
  \lesssim {Ra} \lesssim 2 \times 10^{14}$: ultimate-state transition for
  aspect ratio ${\Gamma} = 1.00$}.  \jt{New J. Phys.}  \bvol{14},  \pg{063030}.

\bibitem[He {\em et~al.\/}(2012{\natexlab{{\em b\/}}})He, Funfschilling,
  Nobach, Bodenschatz \& Ahlers]{he2012b}
{\sc \au{He, X.}, \au{Funfschilling, D.}, \au{Nobach, H.}, \au{Bodenschatz, E.}
  \& \au{Ahlers, G.}} \yr{2012{\natexlab{{\em b\/}}}}  \at{Transition to the
  ultimate state of turbulent {Rayleigh-B\'{e}nard} convection}.  \jt{Phys.
  Rev. Lett.}  \bvol{108},  \pg{024502}.

\bibitem[Huisman {\em et~al.\/}(2012)Huisman, van Gils, Grossmann, Sun \&
  Lohse]{huisman2012}
{\sc \au{Huisman, S.~G.}, \au{van Gils, D. P.~M.}, \au{Grossmann, S.}, \au{Sun,
  C.} \& \au{Lohse, D.}} \yr{2012}  \at{Ultimate turbulent {Taylor-Couette}
  flow}.  \jt{Phys. Rev. Lett.}  \bvol{108},  \pg{024501}.

\bibitem[Huisman {\em et~al.\/}(2013)Huisman, Scharnowski, Cierpka, K\"{a}hler,
  Lohse \& Sun]{huisman2013}
{\sc \au{Huisman, S.~G.}, \au{Scharnowski, S.}, \au{Cierpka, C.},
  \au{K\"{a}hler, C.~J.}, \au{Lohse, D.} \& \au{Sun, C.}} \yr{2013}
  \at{Logarithmic boundary layers in strong {Taylor-Couette} turbulence}.
  \jt{Phys. Rev. Lett.}  \bvol{110},  \pg{264501}.

\bibitem[Ikeda \& Durbin(2007)]{ikeda2007}
{\sc \au{Ikeda, T.} \& \au{Durbin, P.~A.}} \yr{2007}  \at{Direct simulations of
  a rough-wall channel flow}.  \jt{J. Fluid. Mech.}  \bvol{571},  \pg{235}.

\bibitem[Kraichnan(1962)]{kraichnan1962}
{\sc \au{Kraichnan, R.~H.}} \yr{1962}  \at{Turbulent thermal convection at
  arbritrary {Prandtl} number}.  \jt{Phys. Fluids}  \bvol{5},  \pg{1374--1389}.

\bibitem[Lathrop {\em et~al.\/}(1992)Lathrop, Fineberg \&
  Swinney]{lathrop1992b}
{\sc \au{Lathrop, D.~P.}, \au{Fineberg, J.} \& \au{Swinney, H.~L.}} \yr{1992}
  \at{Turbulent flow between concentric rotating cylinders at large reynolds
  number}.  \jt{Phys. Rev. Lett.}  \bvol{68},  \pg{1515--1518}.

\bibitem[Lewis \& Swinney(1999)]{lewis1999}
{\sc \au{Lewis, G.~S.} \& \au{Swinney, H.~L.}} \yr{1999}  \at{Velocity
  structure functions, scaling, and transitions in {high-Reynolds-number}
  {Couette-Taylor} flow}.  \jt{Phys. Rev. E}  \bvol{59},  \pg{5457--5467}.

\bibitem[McKeon {\em et~al.\/}(2005)McKeon, Zagarola \& Smits]{mckeon2005}
{\sc \au{McKeon, B.~J.}, \au{Zagarola, M.~V.} \& \au{Smits, A.~J.}} \yr{2005}
  \at{A new friction factor relationship for fully developed pipe flow}.
  \jt{J. Fluid Mech.}  \bvol{538},  \pg{429--443}.

\bibitem[Nikuradse(1933)]{nikuradse1933}
{\sc \au{Nikuradse, J.}} \yr{1933}  \at{{Str\"{o}mungsgesetze} in rauhen
  {Rohren}}.  \jt{Forschungsheft Arb. Ing.-Wes.}  \bvol{361}.

\bibitem[Ostilla {\em et~al.\/}(2013)Ostilla, Stevens, Grossmann, Verzicco \&
  Lohse]{ostilla2013}
{\sc \au{Ostilla, R.}, \au{Stevens, R. J. A.~M.}, \au{Grossmann, S.},
  \au{Verzicco, R.} \& \au{Lohse, D.}} \yr{2013}  \at{Optimal {Taylor-Couette}
  flow: direct numerical simulations}.  \jt{J. Fluid Mech.}  \bvol{719},
  \pg{14--46}.

\bibitem[Ostilla-M\'{o}nico {\em et~al.\/}(2014{\natexlab{{\em
  a\/}}})Ostilla-M\'{o}nico, van~der Poel, Verzicco, Grossmann \&
  Lohse]{ostilla2014a}
{\sc \au{Ostilla-M\'{o}nico, R.}, \au{van~der Poel, E.~P.}, \au{Verzicco, R.},
  \au{Grossmann, S.} \& \au{Lohse, D.}} \yr{2014{\natexlab{{\em a\/}}}}
  \at{Boundary layer dynamics at the transition between the classical and the
  ultimate regime of {Taylor-Couette} flow}.  \jt{Phys. Fluids}  \bvol{26},
  \pg{015114}.

\bibitem[Ostilla-M\'{o}nico {\em et~al.\/}(2014{\natexlab{{\em
  b\/}}})Ostilla-M\'{o}nico, van~der Poel, Verzicco, Grossmann \&
  Lohse]{ostilla2014b}
{\sc \au{Ostilla-M\'{o}nico, R.}, \au{van~der Poel, E.~P.}, \au{Verzicco, R.},
  \au{Grossmann, S.} \& \au{Lohse, D.}} \yr{2014{\natexlab{{\em b\/}}}}
  \at{Phase diagram of turbulent {Taylor-Couette} flow}.  \jt{J. Fluid Mech.}
  \bvol{761},  \pg{1--26}.

\bibitem[Ostilla-M\'onico {\em et~al.\/}(2016)Ostilla-M\'onico, Verzicco,
  Grossmann \& Lohse]{ostilla2016}
{\sc \au{Ostilla-M\'onico, R.}, \au{Verzicco, R.}, \au{Grossmann, S.} \&
  \au{Lohse, D.}} \yr{2016}  \at{The near-wall region of highly turbulent
  {Taylor-Couette} flow}.  \jt{J. Fluid Mech.}  \bvol{788},  \pg{95--117}.

\bibitem[van~der Poel {\em et~al.\/}(2015)van~der Poel, Ostilla-M\'{o}nico,
  Donners \& Verzicco]{vandepoel2014}
{\sc \au{van~der Poel, E.~P.}, \au{Ostilla-M\'{o}nico, R.}, \au{Donners, J.} \&
  \au{Verzicco, R.}} \yr{2015}  \at{A pencil distributed code for simulating
  wall-bounded turbulent convection}.  \jt{Comput. Fluids}  \bvol{116},
  \pg{10--16}.

\bibitem[Pope(2000)]{pope_book}
{\sc \au{Pope, S.~B.}} \at{ \yr{2000} } \bt{In {\em {Turbulent Flows}\/}}.
  \publ{Cambridge University Press}.

\bibitem[Roche {\em et~al.\/}(2001)Roche, Castaing, Chabaud \&
  H\'{e}bral]{roche2001}
{\sc \au{Roche, P.-E.}, \au{Castaing, B.}, \au{Chabaud, B.} \& \au{H\'{e}bral,
  B.}} \yr{2001}  \at{Observation of the 1/2 power law in {Rayleigh-B\'{e}nard}
  convection}.  \jt{Phys. Rev. E}  \bvol{63},  \pg{045303(R)}.

\bibitem[Schlichting(1979)]{schlichting_book}
{\sc \au{Schlichting, H.}} \at{ \yr{1979} } \bt{In {\em {Boundary layer Theory
  (7th ed.)}\/}}.  \publ{McGraw-Hill}.

\bibitem[Shen {\em et~al.\/}(1996)Shen, Tong \& Xia]{shen1996}
{\sc \au{Shen, Y.}, \au{Tong, P.} \& \au{Xia, K.-Q.}} \yr{1996}  \at{Turbulent
  convection over rough surfaces}.  \jt{Phys. Rev. Lett.}  \bvol{76},
  \pg{908--911}.

\bibitem[Stringano {\em et~al.\/}(2006)Stringano, Pascazio \&
  Verzicco]{stringano2006}
{\sc \au{Stringano, G.}, \au{Pascazio, G.} \& \au{Verzicco, R.}} \yr{2006}
  \at{Turbulent thermal convection over grooved plates}.  \jt{J. Fluid Mech.}
  \bvol{557},  \pg{307--336}.

\bibitem[Tisserand {\em et~al.\/}(2011)Tisserand, Creyssels, Gasteuil, Pabiou,
  Gibert, Castaing \& Chill\`{a}]{tisserand2011}
{\sc \au{Tisserand, J.~C.}, \au{Creyssels, M.}, \au{Gasteuil, Y.}, \au{Pabiou,
  H.}, \au{Gibert, M.}, \au{Castaing, B.} \& \au{Chill\`{a}, F.}} \yr{2011}
  \at{Comparison between rough and smooth plates within the same
  {Rayleigh-B\'{e}nard} cell}.  \jt{Phys. Fluids}  \bvol{23},  \pg{015105}.

\bibitem[Verzicco \& Orlandi(1996)]{verzicco1996}
{\sc \au{Verzicco, R.} \& \au{Orlandi, P.}} \yr{1996}  \at{A finite-difference
  scheme for three-dimensional incompressible flows in cylindrical
  coordinates}.  \jt{J. Comput. Phys.}  \bvol{123},  \pg{402--413}.

\bibitem[Wagner \& Shishkina(2014)]{wagner2014}
{\sc \au{Wagner, S.} \& \au{Shishkina, O.}} \yr{2014}  \at{Heat flux
  enhancement by regular surface roughness in turbulent thermal convection}.
  \jt{J. Fluid Mech.}  \bvol{763},  \pg{109--135}.

\bibitem[Wei {\em et~al.\/}(2014)Wei, Chan, Ni, Zhao \& Xia]{wei2014}
{\sc \au{Wei, P.}, \au{Chan, T.-S.}, \au{Ni, R.}, \au{Zhao, X.-Z.} \& \au{Xia,
  K.-Q.}} \yr{2014}  \at{Heat transport properties of plates with smooth and
  rough surfaces in turbulent thermal convection}.  \jt{J. Fluid Mech.}
  \bvol{740},  \pg{28--46}.

\bibitem[Zanoun {\em et~al.\/}(2009)Zanoun, Nagib \& Durst]{zanoun2009}
{\sc \au{Zanoun, E-S}, \au{Nagib, H} \& \au{Durst, F}} \yr{2009}  \at{Refined
  $c_f$ relation for turbulent channels and consequences for {high- Re}
  experiments}.  \jt{Fluid Dyn. Res.}  \bvol{41}~(2),  \pg{021405}.

\end{thebibliography}

\end{document}